\documentclass[conference]{IEEEtran}
\usepackage{amsmath,amssymb,amsfonts}
\usepackage{caption}
\usepackage{subcaption}
\usepackage{algorithmic}
\usepackage{graphicx}
\usepackage{textcomp}
\usepackage{xcolor}
\def\BibTeX{{\rm B\kern-.05em{\sc i\kern-.025em b}\kern-.08em
    T\kern-.1667em\lower.7ex\hbox{E}\kern-.125emX}}
\usepackage[backend=biber,
            style=numeric, 
            abbreviate=false,
            dateabbrev=false,
            alldates=long,
            maxbibnames=6,
            giveninits=true
            ]{biblatex}
\usepackage{cuted}
\usepackage{soul}
\usepackage{svg}
\usepackage{dblfloatfix}
\usepackage{float}
\usepackage{hyperref}
\AtBeginBibliography{\footnotesize}
\usepackage{hyperref}
\newcommand\rurl[1]{%
  \href{http://#1}{\nolinkurl{#1}}%
}

\begin{document}

\title{RetinaNet Object Detector based on Analog-to-Spiking Neural Network Conversion\\
}


\author{\IEEEauthorblockN{Joaquín Royo Miquel\IEEEauthorrefmark{1}, Silvia Tolu\IEEEauthorrefmark{2}, 
Frederik E. T. Schöller\IEEEauthorrefmark{3},  and
Roberto Galeazzi\IEEEauthorrefmark{4}}
\IEEEauthorblockA{Technical University of Denmark, Department of Electrical Engineering, Automation and Control Group. \\
DK 2800 Lyngby, Denmark\\
Email: \IEEEauthorrefmark{1}joaromi@outlook.com,
\IEEEauthorrefmark{2}stolu@elektro.dtu.dk,
\IEEEauthorrefmark{3}fets@elektro.dtu.dk,
\IEEEauthorrefmark{4}rg@elektro.dtu.dk}}

\maketitle

\begin{abstract}
The paper proposes a method to convert a deep learning object detector into an equivalent spiking neural network. The aim is to provide a conversion framework that is not constrained to shallow network structures and classification problems as in state-of-the-art conversion libraries. The results show that models of higher complexity, such as the RetinaNet object detector, can be converted with limited loss in performance.
\end{abstract}


\section{Introduction} \label{Introduction}
Over the past few decades, artificial neural networks have become the new standard in artificial intelligence (AI) and top performers in increasingly complex tasks. Although conceptually they bear a strong biological inspiration, trade-offs have been made for their execution in modern CPU/GPU architectures that dissociate them from their natural roots in the animal brain.

This issue has been recently addressed by the emergence of neuromorphic platforms, i.e. hardware that electronically recreates the interconnections and dynamics of the nervous system for the instantiation of bio-plausible, spiking neural networks (SNNs). There are high expectations that these \textit{Silicon Brains} will be more computationally efficient while coming closer to natural intelligence than former generations of neural networks \cite{silvia}. However, such a shift of paradigm entails a large breach with these previous AI frameworks, which hinders the use of the more widespread learning techniques and limits the current applications of neuromorphic computing.

With the aim of bridging this gap, research is being made on conversion algorithms \cite{ANN2SNN_2013, ANN2SNN_2015_diehl} which enable the porting of trained deep-learning models to SNNs minimizing the performance loss. While still very embryonic, these techniques have been applied on shallow models achieving remarkable performances in classification tasks \cite{ANN2tSNN} and more recently, achieved a spiking version of the Tiny-YOLO object detector running on the \textit{TrueNorth} neuromorphic chip \cite{Spiking-YOLO}.

This paper, details how the RetinaNet convolutional neural network (CNN) object detector \cite{RetinaNet} has been implemented as a SNN and tested in simulation based on the conversion algorithm in \cite{ANN2rSNN}. The mentioned algorithm, which was originally conceived for shallow classification experiments, has been enhanced to fit the higher complexity of RetinaNet making it suitable for object detection problems. The main changes have been:
\begin{itemize}
    \item The adaptation of the pipeline to a multi-network input model and the temporal-mean-rate (spike) encoding of the object detection output.
    \item An adapted normalization method to enable the use of simple \textit{Integrate-and-Fire} (IF) neurons to build the Spiking-RetinaNet.
\end{itemize} 

The document is structured to provide some background knowledge in section \ref{Relevant Concepts}, explaining afterwards the conversion algorithm in section \ref{conversion_method}. The obtained results on the RetinaNet model are evaluated in section \ref{Results} and discussed in section \ref{Discussion}. Section \ref{Conclusion} draws some conclusions on the research work.

The code used in the project is available at: {\small\texttt{\rurl{github.com/joaromi/Spiking-RetinaNet}}}.


\section{Relevant Concepts} \label{Relevant Concepts}

\subsection{Spiking Neural Networks}

Artificial spiking neural networks are deep learning models that closely mimic the behavior of biological NNs \cite{silvia2}. They are identified by their inner communications, driven by sparse, asynchronous binary signals produced by the neuronal dynamics. Also referred to as third generation neural networks, they were initially approached for neuroscientific purposes. Nevertheless, SNNs exhibit favorable attributes for deep-learning such as fast inference, distributed computing and low-power consumption due to event-driven information processing \cite{SNN_overview}.

SNNs can be instantiated within a software simulation or in neuromorphic processing units, where they display their full potential. Some examples of these hardware platforms are the University of Manchester's \textit{SpiNNaker}, IBM's \textit{TrueNorth} and Stanford's \textit{Neurogrid}.

As opposed to SNNs, conventional artificial neural networks with static neuron models will be referred as analog neural networks (ANNs) following the nomenclature in \cite{ANN2rSNN}.

\subsection{RetinaNet}
RetinaNet \cite{RetinaNet} is a state-of-the-art one-stage object detector that achieves performances comparable to classical two-stage approaches (e.g. Faster R-CNN) while being computationally cheaper. The key to these results is the novel loss function it incorporates, named \textit{Focal Loss}, which addresses the imbalance between the trivial but over-represented background class and the object classes. It is a single, unified convolutional neural network consisting of a feature detector backbone (\textit{Feature Pyramid Network}) and two task-specific sub-networks in charge of the bounding box regression and object classification.

\begin{figure}[!b]
    \centering
    \begin{subfigure}{1\linewidth}
    \includegraphics[width=\linewidth]{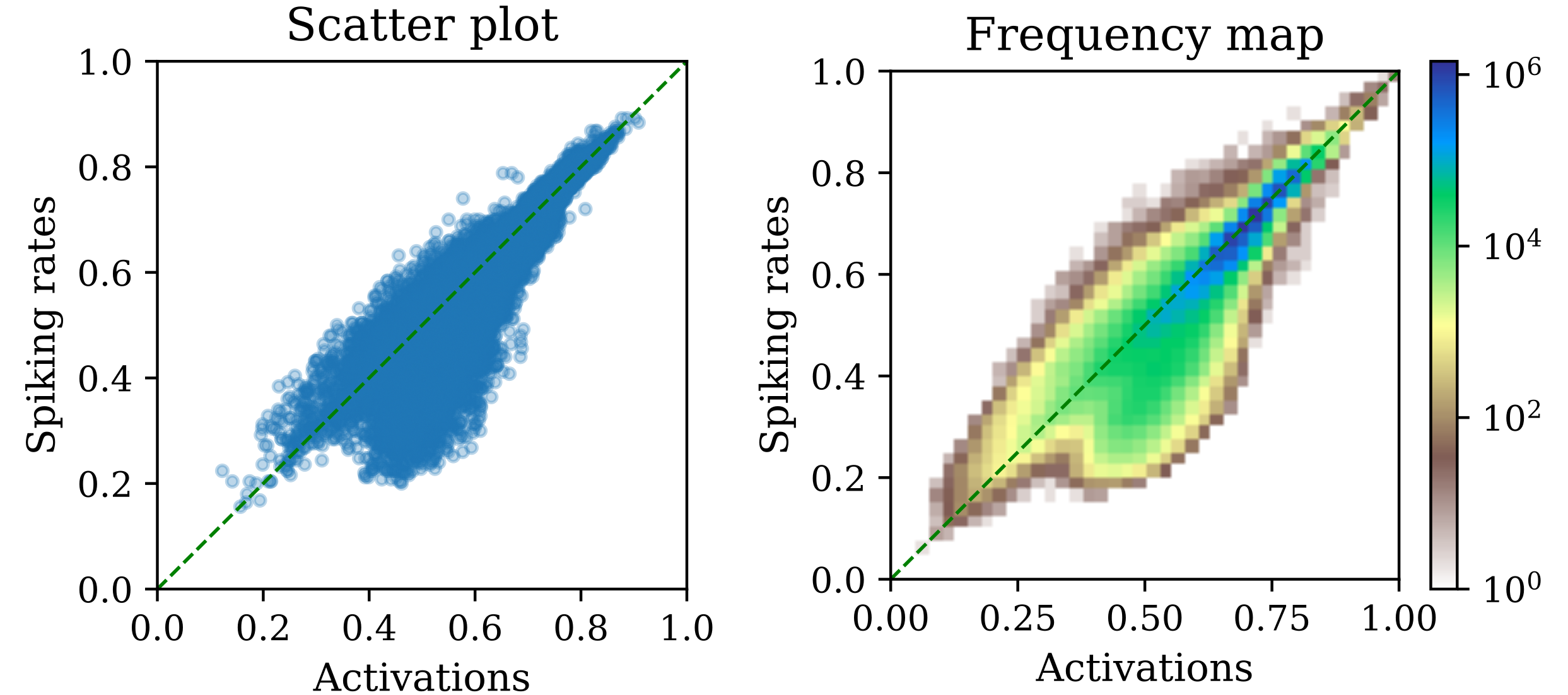}
    \caption{$500ms$ simulated time window.}
    \end{subfigure}
    \begin{subfigure}{1\linewidth}
    \includegraphics[width=\linewidth]{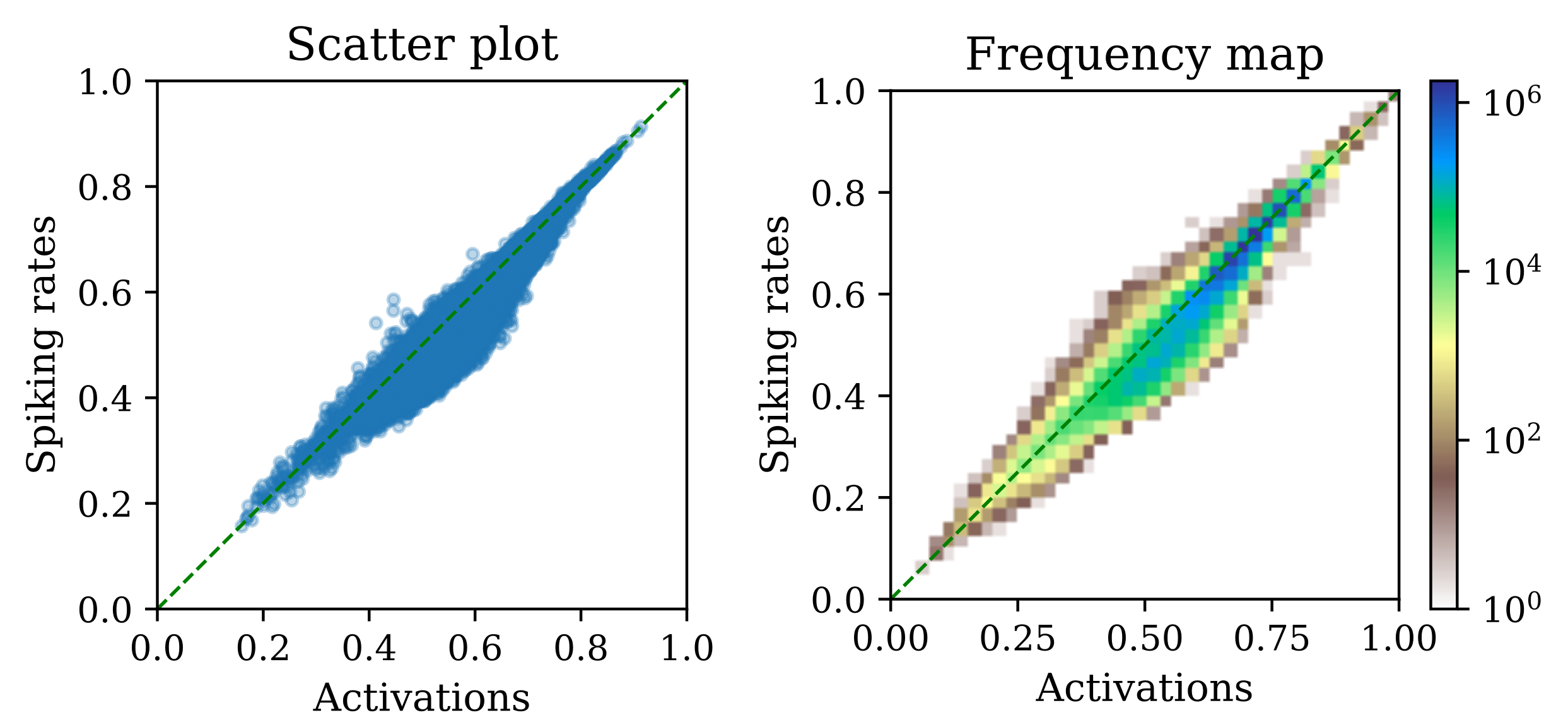}
    \caption{$2000ms$ simulated time window.}
    \end{subfigure}
    \caption{Correlations between the original analog activations and obtained spiking rates after the ANN-to-SNN conversion of RetinaNet. The data corresponds to the output layer. Notice the convergence over the model run-time when comparing the correlations at $500ms$ (a) and $2000ms$ (b) timestamps.}
    \label{FIG_correlations}
\end{figure}

\subsection{ANN-to-SNN conversion} 
Despite their excellent potential, SNN applications are currently limited by the lack of scalability of their learning algorithms, demanding research on more suitable techniques. This is caused by the complex dynamics and non-differentiable activations of spiking neurons. Progress has recently been made on parallel conversion approaches, which rely on the idea of translating a pre-trained ANN to an equivalent SNN with minimal sacrifice in performance.

Early studies on this topic were initiated in 2013 with the research performed by Perez-Carrasco et al. \cite{ANN2SNN_2013}. A close link between the dynamics of a spiking neuron and a ReLU activation function was suggested by Cao et al. \cite{ANN2SNN_2015_cao} and improved by Diehl et al. \cite{ANN2SNN_2015_diehl}, achieving nearly loss-free conversion on the MNIST dataset using a parameter normalization step. Directly serving as the foundation for this project, the work of Rueckauer et al. \cite{ANN2rSNN} was built upon previous works and gave support for many operators that are crucial for improved ANN error scores. The underlying encoding scheme for this conversion approach is rate-based (r-SNN), i.e. the neurons generate spike-trains whose rate approximates the analog activations corresponding to the original ANN. This encoding becomes more accurate as the simulation duration is increased entailing a higher resolution at the price of scaling up the computational cost.

In 2019 Kim et al. \cite{Spiking-YOLO} implemented the first object detection r-SNN, \textit{Spiking-YOLO}, in neuromorphic hardware. The model achieved comparable results to the original ANN in non-trivial datasets with 2\% performance reduction and 280 times less energy consumption. This was achieved through a fine-grained normalization method that raised the information transmission, and by the design of a new spiking neuron model, able to transmit negative values using an imbalanced threshold.

\section{Conversion method} \label{conversion_method}

\begin{figure*}[tbp]
    \centering
    \includegraphics[width=1\textwidth]{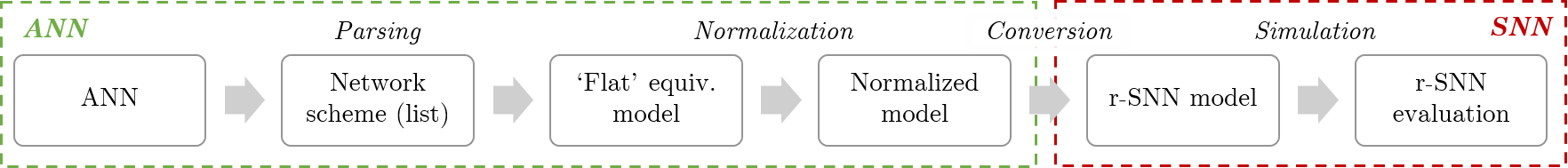} 
    \caption{ANN-to-SNN conversion pipeline structure, as first used in \cite{ANN2rSNN} and preserved in our implementation.}
    \label{FIG_pipeline}
\end{figure*}

The basis of the method follows the algorithm proposed by Rueckauer et al. in \cite{ANN2rSNN} and its implementation in \cite{SNN_Toolbox}. The conversion pipeline (fig. \ref{FIG_pipeline}) has, however, received several updates to accommodate the structure and syntax of object detection convolutional neural networks (CNNs).

\subsection{Rate-encoding of the activations}

The method used for the ANN to r-SNN conversion is based on the principle that the firing rate of the spiking neurons over a certain time window should approximate the activations of the original analog neurons \cite{ANN2rSNN}. A one-to-one correspondence between analog and spiking neurons is assumed.

The dynamics of the neurons in the SNN are modelled using the \textit{Integrate and Fire (IF)} membrane equations. The $i$-th neuron in the $l$-th network's layer has a membrane potential $V_i^l(t)$, which is driven by the input current $z_i^l(t)$ and the event of the generation of a spike $\Theta_i^l(t)$. 
\begin{align}
V_{i}^{l}(t)&=V_{i}^{l}(t-1)+\underbrace{z_{i}^{l}(t)}_{\text {input }}-\underbrace{V_{th} \Theta_{i}^{l}(t-1)}_{\text {reset }} 
\\
z_{i}^{l}(t)&=V_{th}\left(\sum_{j=1}^{M^{l-1}} W_{i j}^{l} \Theta_{j}^{l-1}(t)+b_{i}^{l}\right)
\\
\Theta_{i}^{l}(t)&=\left\{\begin{array}{ll}
1 & \text { if } V_{i}^{l}(t) \geq V_{th} \quad\longleftarrow \text{\footnotesize \textit{ spike}}\\
0 & \text { else. }
\end{array}\right.
\end{align}
The parameter $V_{th}$ is the membrane voltage threshold that triggers the spiking $\Theta_{i}^{l}(t)$ and posterior reset of the neuron (\textit{reset by subtraction}). The input current $z_i^l(t)$ is obtained as a linear combination of the spikes $\Theta_{j}^{l-1}(t)$ from the previous layer.

The \textit{IF} neuron firing rate is assumed to range from the non-spiking state of the neuron (0\%) to its constant firing state (100\%). Achieving a one-to-one correspondence of the spike rates to the original activations requires these to fall in the unit interval, which is accomplished through a prior normalization of the ANN weights. The SNN is then built as an equivalent copy of the structure and parameters of the ANN. 

The accuracy of the conversion is best measured by studying the correlations between each layer's activations in the original ANN and the spike rates of the generated SNN (fig. \ref{FIG_correlations}). Two main sources of error can be identified \cite{ANN2rSNN_theory}: 
\begin{itemize}
    \item Neuron firing saturation for activations that fall outside the 0$-$1 range.
    \item Time-decaying error caused by the discrete-sampling of the analog activations in the spike trains. This is the reason why correlations in fig. \ref{FIG_correlations} improve over time.
\end{itemize}

These two error sources are addressed in the normalization phase prior to conversion. However, there is a trade-off to be made between them. Prioritizing the complete removal of saturation means fitting all activations in the unit interval, which may produce very low firing rates in a significant percentage of the neurons. This will increase the simulation time needed to reduce the discrete sampling error. As previously done by Rueckauer et al. \cite{ANN2rSNN}, the normalization will instead fit 99.99\% of the activations. Good empirical results were obtained with this amount.

\subsection{Conversion pipeline}

The conversion algorithm follows the scheme displayed in fig. \ref{FIG_pipeline}. The pipeline, built using the Keras framework, takes in an ANN model and produces a functional equivalent spiking neural network. It is divided in four very distinct stages which are detailed below.

\vspace{1mm}
\subsubsection{Parsing}

To handle input model variability, the toolbox uses a first translation step. Relevant information is extracted from the input ANN and processed to create a dictionary that makes this data immediately accessible. An equivalent (parsed) model is then tailored for an optimal ANN-to-SNN porting.

The parsing step handles different types of layers or, as a new feature, sub-networks in the input model, for posterior porting to the spiking model.

\begin{figure}[b!]
\centering
\centerline{\includegraphics[width=\linewidth]{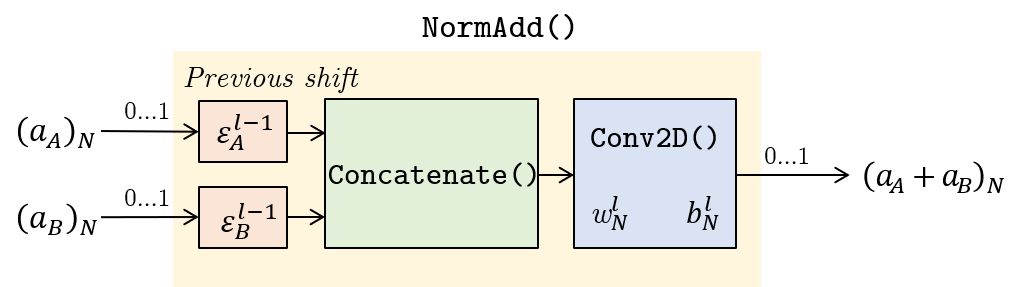}}
\caption{Structure of the \textit{NormAdd} layer. It is designed to preserve the relative scales and locations of the input values prior to the summation, so that the information is correctly transmitted through the spiking layer.}
\label{normadd}
\end{figure}

\vspace{1mm}
\subsubsection{Normalization}

The normalization step receives the parsed model and modifies the parameters of its layers so that every activation falls within the 0$-$1 interval while preserving their logic.

The original layer normalization method used in \cite{ANN2rSNN} has two issues: it leads to low firing rates and does not account for negative activation values. In \cite{Spiking-YOLO}, the first issue is significantly improved by performing the normalization channel-wise instead of layer-wise. This approach magnifies channels with small activations and increases the firing rates of the network. 

To address the conversion of negatively-activated neurons to IF spiking neurons, a novel normalization method has been introduced which builds on top of the latter. The method uses an estimate of activation values distribution in each channel to fit them in the unit range through a shift and a scaling. This estimate consists of the 0.01 and 99.99th percentiles of the activation values, computed channel-wise over a significant portion of the training data ($\sim10\%$). These values will be referred to as $\varepsilon$ and $\lambda$.

The desired behavior of the network after normalization can be better explained as a multi-step recursion. The normalized activation of channel $i$ in layer $l$ ($\tilde{a}_{i}^l$) relates to the original activation ($a_i^l$) as
$\tilde{a}_{i}^l = (a_i^l-\varepsilon_i^l) (\lambda_i^l-\varepsilon_i^l)^{-1}$.
A downstream neuron in layer $l+1$ will get this normalized activation, decode it to its original magnitude, compute its own activation, and finally normalize it.

This process, however, is performed in a single step and absorbed to the convolution weights of each channel as shown in \eqref{SNN_norm_eqns}, where $\tilde{w}_{i, j}^{l}$ is the normalized kernel from channel $i$ in layer $l-1$ to channel $j$ in layer $l$, and $\tilde{b}_{j}^{l}$ is the normalized bias for channel $j$ in layer $l$.
\begin{align}  \label{SNN_norm_eqns}
\text{I}& \text{nput layers:}           
&  
\text{H}& \text{idden layers:}
 \notag \\
\tilde{w}_{i, j}^{l}& = w_{i, j}^{l} \frac{1}{\lambda_{j}^{l}-\varepsilon_{j}^{l}}       
&
\tilde{w}_{i, j}^{l}& = w_{i, j}^{l} \frac{\lambda_{i}^{l-1}-\varepsilon_{i}^{l-1}}{\lambda_{j}^{l}-\varepsilon_{j}^{l}}
\\
\tilde{b}_{j}^{l}&=\frac{b_{j}^{l}-\varepsilon_{j}^{l}}{\lambda_{j}^{l}-\varepsilon_{j}^{l}}         
&
\tilde{b}_{j}^{l}&=\frac{\left[b_{j}^{l} + \sum_i\left(w_{i, j}^{l}\varepsilon_{i}^{l-1}\right)\right]
-\varepsilon_{j}^{l}}{\lambda_{j}^{l}-\varepsilon_{j}^{l}} 
 \notag
\end{align}

\begin{figure}[b]
\centering
\centerline{\includegraphics[width=\linewidth]{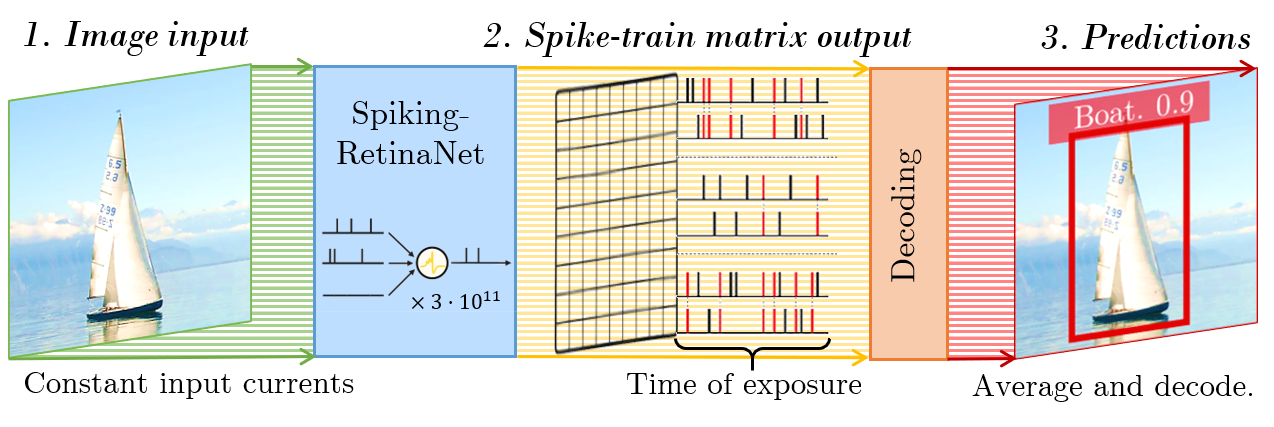}}
\caption{Illustrative scheme of the r-SNN simulation for a specific time window.\\}
\label{FIG_sim}
\end{figure}

\begin{figure*}[t]
    \centering
    \begin{subfigure}{0.49\textwidth}
        \includegraphics[width=\textwidth]{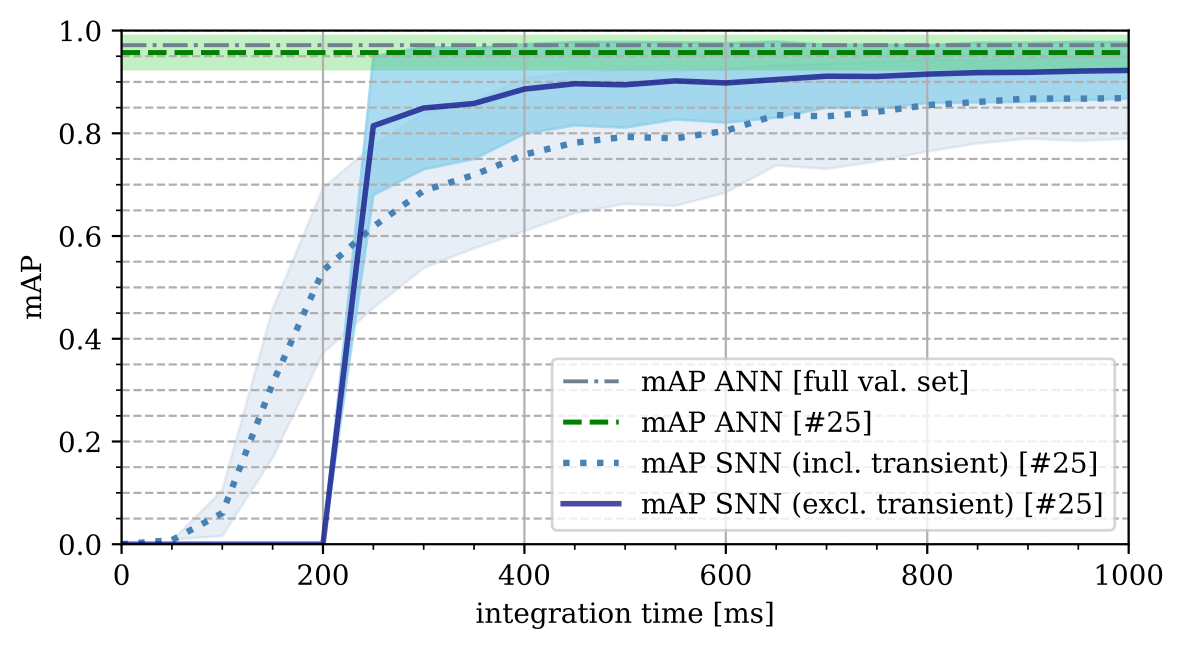}
        \caption{mAP@0.5IoU score on ShippingLab subset (2 classes)}
        \label{FIG_mAP_sl}
    \end{subfigure}
    \begin{subfigure}{0.49\textwidth}
        \includegraphics[width=\textwidth]{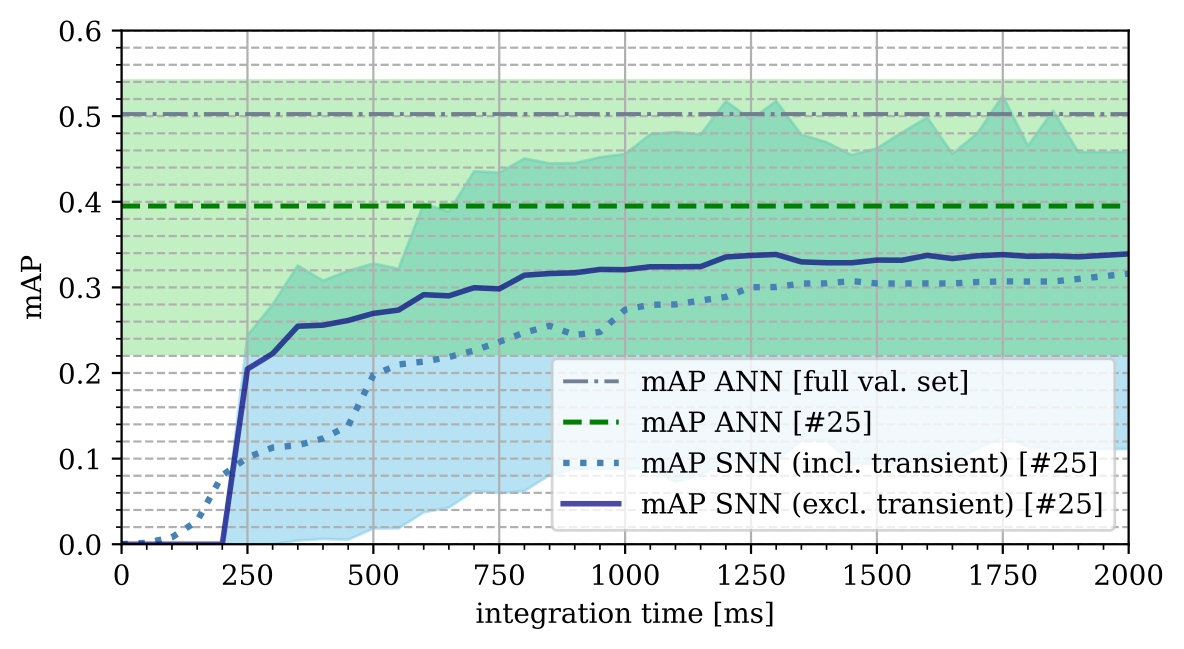}
        \caption{mAP@0.5IoU score on MS COCO subset (80 classes)}
        \label{FIG_mAP_coco}
    \end{subfigure}
    \caption{Mean Average Precision evolution with simulation time of the spiking-RetinaNet model (computed every $50ms$). Due to the simulation's computational constraints, the analysis was performed using 25 relevant samples from each the ShippingLab (a) and COCO (b) validation sets. The original RetinaNet model score is displayed as a reference (green dashed line).}
    \label{FIG_mAP}
\end{figure*}

As a downside of this normalization method, spiking layers with multiple inputs need to be redesigned to correctly preserve their logic after the SNN conversion. In the case of RetinaNet, \textit{Add} layers had to be substituted by the \textit{NormAdd} layer shown in fig. \ref{normadd}.

\vspace{1mm}
\subsubsection{Conversion}
The normalized model is now mirrored using equivalent spiking counterparts for all its neurons. These layers were previously implemented in the Keras-based INI simulator using temporal-mean-rate encoding and the IF neuron model. All bias parameters are adjusted to the simulation time step as $\hat{b}_i^l = {b}_i^l \cdot \Delta t$, as suggested in \cite{ANN2rSNN}.

\vspace{1mm}
\subsubsection{Simulation}
The obtained spiking object detector is exposed to the input image within a given time window (see fig. \ref{FIG_sim}). The image is fed to the model as a matrix of constant currents and propagates through the network driving the final layer to a steady spiking rate. After these rate values are denormalized they can be processed and decoded in the same way as the original ANN output activations.

\section{Results} \label{Results}

The implementation and testing of this conversion method on an object detector has been framed within the ShippingLab \cite{shippinglab} research initiative towards the development of a perception system for autonomous navigation of surface vessels. The main analysis was therefore performed on the ShippingLab marine dataset, a pool of images from different sea scenarios that are are annotated in two main classes: \textit{boats} (sea vehicles) and \textit{buoys} (stationary obstacles). Subsequently, and with the aim of benchmarking the method on a standardized and more challenging environment, a second evaluation was carried out using the MS COCO dataset \cite{MS_COCO}.

It is important to clarify that the spiking neural networks’ results have been obtained by emulation in the INI/Keras framework from \cite{SNN_Toolbox}, with a relatively high computational cost. The objective of this framework is to act as a translation and validation step towards the future porting of the obtained SNN to a neuromorphic platform, where it would run in real time and benefit from its low energy footprint.

The original (ANN) RetinaNet models used for the conversion achieve a mAP@0.5IoU\footnote{A mean Average Precision (mAP) with an Intersection over Union (IoU) of 0.5 for true positive detections.} of 50.25\% on the COCO data and 97.16\% on the ShippingLab data.


As shown in fig. \ref{FIG_mAP} the performance of the obtained spiking models converges to the one achieved by the analog network given enough running time to counteract the approximation errors. The evolution of these can be appreciated by checking the correlations between the original activations and the obtained spiking rates for the output layer of RetinaNet (fig. \ref{FIG_correlations}). With a larger time window the spike trains exhibit a higher resolution, translating into a lower discrete sampling error and more accurate predictions.

An improvement in convergence time has been reached by ignoring spikes produced in the initial transient state of the network, which is the time needed for the spiking rate of the last layers to stabilize. The delay in the propagation of the information through such a deep spiking model causes these first spikes to add a substantial error to the calculated rates. While it will eventually be compensated by future spikes in the simulation, this will be time-consuming. By modifying the prediction method to ignore outputs in this time interval it is possible to functionally speed up the convergence of the predictions between 3 and 5 times.

The score of the SNN converges to the ANN performance in the ShippingLab data set with a loss of 2\% in $1000ms$ (fig. \ref{FIG_mAP_sl}) achieving a mAP@0.5IoU of 93.43\%. Regarding the COCO data set, its bigger class collection demands a much larger resolution to encode the predictions. The loss in performance obtained after $1250ms$ is of roughly 12\% (fig. \ref{FIG_mAP_coco}), a value that could be improved with a longer time of exposure in applications where detection speed is not a decisive factor. The obtained mAP@0.5IoU is of 34.85\%.

It should be acknowledged that, due to the computational cost of the simulation, the mAP test shown on fig. \ref{FIG_mAP} was performed over a sample of 25 images for each of the datasets. The samples were picked to contain a good representation of all classes and are liable to provide a good preliminary insight into the real performance of the spiking model. However, the obtained ANN mAP scores are lower than with the full validation sets (96.01\% for ShippingLab and 39.13\% for COCO). Analysis on a larger dataset should be conducted when implemented in neuromorphic hardware to check the correctness of these first results.

\section{Discussion} \label{Discussion}

A few trade-offs made in this SNN object detector implementation should be highlighted:

Regarding the chosen encoding scheme for the spike-trains, achieving state-of-the art accuracy with rate-based networks comes at the cost of using high firing rates and long integration times to obtain reliable results. This reduces the energy consumption gap between traditional methods and SNNs. Using temporal encoding is an attractive alternative, but, for the moment, these approaches could not be easily adapted to the object detection problem.

The use of basic \textit{Integrate-and-Fire} neurons for the porting of the off-the-shelf RetinaNet model to a SNN required a complex normalization method and the custom \textit{NormAdd} layer. This layer slightly increases the intricacy of an already big network and the potential difficulty of implementing it on a neuromorphic platform. Possible alternatives would be the use of a different spiking neuron model or a strictly-positive RetinaNet version as base ANN.

Finally, the main limitation of these ANN-to-SNN conversion methods is that they are subject to the constraints and performance cap of the ANN they take as input. The behavior of the spiking models during training and prediction would also benefit of the use of event-based data as input, as SNN show better results in online learning and precise-timing frameworks \cite{SNN_overview} instead of picture datasets tailored for traditional deep-learning methods.

\section{Conclusions} \label{Conclusion}

In this paper, an existing ANN-to-SNN conversion method \cite{ANN2rSNN} has been upgraded with the aim of running convolutional object detection models in a spiking encoding. In contrast with this previous method, which had only been applied to convert simple networks used for classification tasks, the enhanced conversion mechanism has been proven capable of generating a spiking version of the RetinaNet CNN based on the widespread \textit{Integrate-and-Fire} neuron model. In a simulation environment, the generated SNN has undergone preliminary testing and shown promising results on both the ShippingLab and the COCO data sets. The comparison of performance in terms of convergence rate and loss against the only existing spiking object detector, Spiking-YOLO \cite{Spiking-YOLO}, will demand an implementation of the SNN converted RetinaNet on neuromorphic hardware.

Although these approaches are still immature, ANN-to-SNN conversion increases the versatility of neuromorphic platforms by enabling them to execute traditional deep-learning algorithms in addition to their bio-inspired learning capabilities. We believe this is an important step towards more energy efficient artificial intelligence and the \textit{Internet of Things} (IoT) paradigm, as future work on the neuromorphic hardware implementation of this method could pave the way for low-power, high-accuracy deep-learning.

\section*{Acknowledgments}

The research leading to this paper was conducted within the ShippingLab research program \cite{shippinglab} sponsored by the Danish Innovation Fund, The Danish Maritime Fund, Orients Fund and the Lauritzen Foundation, Grant number 8090-00063B.

\begin{figure}[t!]
\centering
\centerline{\includegraphics[width=1\linewidth]{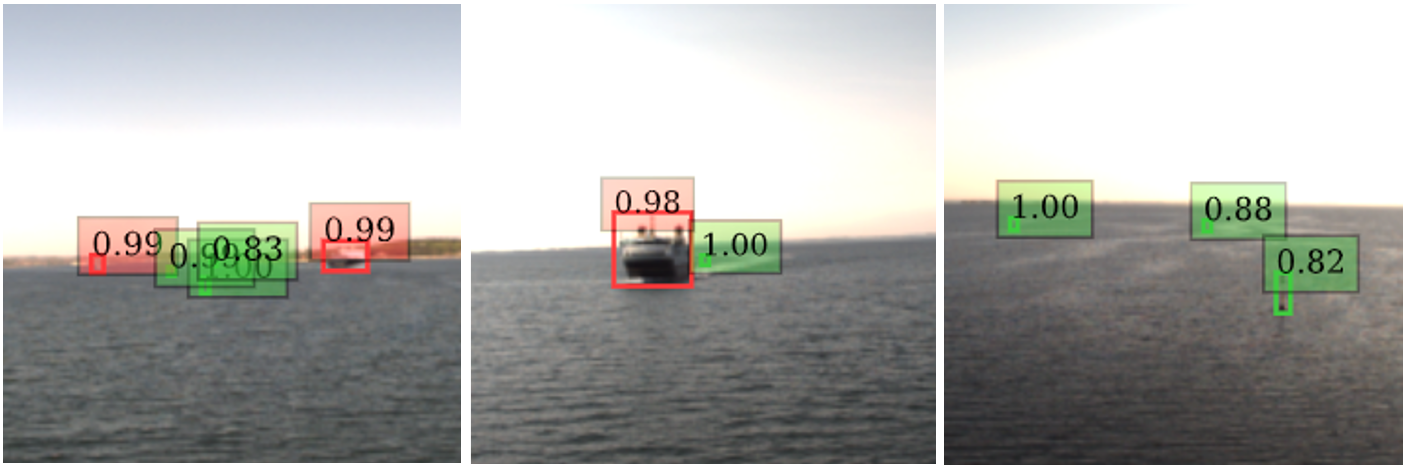}}
\caption{Object detections performed by the spiking-RetinaNet model with a $1000ms$
integration time-window and disregarding of the transient. Boats are framed red and
buoys are framed green; confidence scores are specified. Images belong to the ShippingLab dataset.}
\label{detections}
\end{figure}

\printbibliography


\end{document}